\documentclass[12pt]{iopart}

\usepackage{color}
\usepackage[dcu]{harvard}
\newcommand\bm[1]{\mbox{\boldmath $#1$}}

\def\npk{\bm{k}}
\def\npl{\bm{l}}

\newtheorem{theorem}{Theorem}
\newtheorem{lemma}{Lemma}

\newcommand{\eqref}[1]{(\ref{#1})}

\def\CKP{Cartan-Karlhede procedure}

\begin{document}

\title{Totally symmetrized spinors and null rotation invariance}

\author{ M. A. H. MacCallum\\
School of Mathematical Sciences,\\
Queen Mary University of London, \\ Mile End Road, 
London E1 4NS, UK\\
Email: M.A.H.MacCallum@qmul.ac.uk\\
}
\date{\today}
\maketitle
\begin{abstract}
 In the existing implementations of the \CKP\ for characterization and
 classification of spacetimes, a prominent r\^ole is played by
 multi-index two-component spinors symmetrized over both types of
 index. This paper considers the conditions for, and detection of,
 null rotational invariance of such spinors, and corrects a previous
 discussion.
\end{abstract}

\section{Introduction}

The Cartan-Karlhede procedure for characterizing spacetimes is set out
in several places, e.g.\ \citeasnoun{SteKraMac03}, Chapter 9. It
relies on computing and interpreting Cartan invariants, which are
components of the Riemann tensor and its covariant derivatives
expressed relative to a canonically-chosen frame. To apply this method
the present implementations
\cite{MacSke94,PolSkedIn00,PolSkedIn00a,PolSkedIn00b} use
two-component spinors in the Newman-Penrose formalism (see
\citeasnoun{SteKraMac03}, Chapters 3-7 for a summary of relevant
concepts and formulae).

\citeasnoun{MacAma86} defined a minimal set of Cartan invariants,
taking into account their interrelations, and sufficient for the
\CKP. This set of quantities consists of totally symmetrized spinors.
A shorthand notation for these will be used here: it first appeared in
print in \citeasnoun{KarAma80}.  It is an extension of the
Newman-Penrose notation, in which, for example, the completely
symmetric Weyl spinor $\Psi_{ABCD}=\Psi_{(ABCD)}$ is represented by 5
components $\Psi_{A},~A=0\ldots 4$, formed by contractions with the
basis spinors $(o^B,\,\iota^B)$.  The index $A$ on $\Psi_{A}$ counts
the number of contractions with $\iota^A$: thus, for example,
$\Psi_{3}=\Psi_{(ABCD)}o^A\iota^B\iota^C\iota^D$.

This notation is extended as follows. Suppose
$Q^{BCD\ldots}{}_{E'F'\ldots}$ is a completely symmetric multi-index
spinor, so that
$Q^{BCD\ldots}{}_{E'F'\ldots}=Q^{(BCD\ldots)}{}_{(E'F'\ldots)}$. (Here
the unprimed indices have been raised to enable clarity in the use of
the standard notation for symmetrization).  Such a spinor is said to
have valence $(m,\,n)$ if it has $m$ unprimed and $n$ primed
indices. Then $Q_{ab'}$ denotes the component in which $a$ of the
unprimed indices and $b$ of the primed indices are contracted with the
basis spinors $\iota^A$ and $\bar{\iota}^{W'}$ respectively (and the
other indices with $o^A$ and $\bar{o}^{X'}$).  (Lower case indices
will be used here in $Q_{ab'}$ for consistency with the earlier paper
on the present topic, \citeasnoun{PolSkedIn00}. Normally the same
upper case letters are used for both spinor indices and shorthand
notation indices where no confusion can arise.) As an example,
consider the totally symmetrized part of the second covariant
derivative of the Weyl spinor,
$\nabla_{AA'}\nabla_{BB'}\Psi_{CDEF}$. It has valence $(6,\,2)$. The
component $\nabla^2\Psi_{41'}={\nabla^{\left( A \right.}}_{\left(
  X'\right.}{\nabla^B}_{\left. W'\right)}\Psi^{\left. CDEF\right)}
o_Ao_B\iota_C\iota_D\iota_E\iota_F\bar{o}^{X'}\bar{\iota}^{W'}$.

To identify when a spacetime admits isometries, the software CLASSI
applies a number of tests to the calculated symmetrized spinor Cartan
invariants. These tests are incompletely set out in the published
description \cite{MacSke94} and in the manual \cite{Ama87},
although the code is publically available.  A more extended
description of the process, with (as compared with CLASSI's version)
modified, corrected and extended details, was given by
\citeasnoun{PolSkedIn00}, who also developed its
implementation in Maple.

In their work \citename{PolSkedIn00} gave conditions for symmetric
spinors $\chi_{ab'}$ of valence $(m,n)$ to be invariant under a group
of null rotations preserving the Newman-Penrose basis vector
$\npk$. For invariance under a two-parameter group of null rotations,
they showed that only $\chi_{mn'}$ can be non-zero.

They also gave conditions for invariance under a one-parameter group
of null rotations about $\npk$, for brevity referred to here as IGNR,
specialized to the cases where the parameter $B$ of
\citeasnoun{SteKraMac03}, equation (3.15), [$\alpha$ in
  \citeasnoun{PolSkedIn00}] is either pure real or pure
imaginary. Such a null rotation can be expressed in the spinor basis
as
\begin{equation}\label{nulrot}
  o^A \rightarrow o^A,\qquad \iota^A \rightarrow \iota^A + B o^A,
\end{equation}
for a complex number $B\neq 0$. Note that considering only these null
rotations, rather than the ones preserving $\iota$ and $\npl$, amounts
to preferring, for example, the standard form for the Weyl tensor in
which for Petrov type N only $\Psi_4\neq 0$. In practice an alternate
canonical form with, e.g., only $\Psi_0\neq 0$ may arise or be preferred:
swapping $o$ and $\iota$ has the effect that in the arguments and
statement below $\chi_{ab'}$ is swapped
with $\chi_{(m-a)(n-b)'}$, and $B$ with the parameter $E$ in the null
rotation  $\iota^A \rightarrow \iota^A$, $o^A \rightarrow o^A+E\iota^A$.   

For both $B$ real and $B$ pure imaginary,
\citename{PolSkedIn00} state that one has
\begin{equation}\label{chizero}
  \chi_{ab'}=0, ~~\forall ~{\rm pairs}~ a+b <m+n-2.
\end{equation}
Unfortunately their argument for \eqref{chizero}, and \eqref{chizero}
itself, are false. They overlooked the fact that their equations (11),
(14), and (16)-(19) are not independent. Specifically (18) and (19)
imply (16), (17) and (18) imply (14), and (17)--(19) imply (11). Thus
they have only 3 independent conditions for the four $\chi_{ab'}$,
$a+b=m+n-3$, and setting all those $\chi_{ab'}$ to zero is not the
only solution.

The issue is thus re-examined here, and a correct set of conditions
for totally symmetrized spinors to be invariant under a one-parameter
group of null rotations is given.  Procedures for identifying when
this occurs are suggested.

\section{Conditions for null rotation invariance}

In the following the r\^oles of primed and unprimed indices (and thus
of $m$ and $n$) can be interchanged, so without loss of generality it
can be assumed that $m\geq n$ (this is anyway the case for the
particular spinors defined in the minimal set of \citeasnoun{MacAma86}).
For $m=0=n$ IGNR necessarily applies, so we can assume $m\geq 1$.

First consider $n=0$. Then under \eqref{nulrot}
\begin{equation}\label{mrow}
  {\chi_m}^\star = \sum_{r=0}^m
  \left( \begin{array}{c}m\\r\end{array}\right) B^r \chi_{(m-r)},
\end{equation}
where $\star$ denotes the transformed value, so by equating the
coefficients of this polynomial in $B$ to zero we find that all terms
$\chi_p=0$ for $0\leq p \leq (m-1)$. For IGNR, only $\chi_m$ can be
nonzero.

Now suppose $n \geq 1$. The general formula for the transform
under \eqref{nulrot} of the component $\chi_{ab'}$ of $\chi$ is
\begin{equation}\label{chistar}
  {\chi_{ab'}}^\star = \sum_{r=0}^a \sum_{s=0}^b
  \left( \begin{array}{c}a\\r\end{array}\right)
    \left( \begin{array}{c}b\\s\end{array}\right)
  B^r\bar{B}^s \chi_{(a-r)(b-s)'}.
\end{equation}
The set of values $\chi_{rs'}$ for $r+s=k$ will be referred to as the
line $k$.  For each $k\geq 1$ the terms in \eqref{chistar} homogeneous
of degree $k$ in $B$ and $\bar{B}$, which will be denoted by
$\chi_{ab'}^{(k)\star}$, must vanish in order for $\chi_{ab'}$ to be
IGNR. This can be shown by writing $B=|B|e^{i\theta}$ and equating
coefficients of $|B|^k$ in \eqref{chistar} to zero. If $\chi_{ab'}$
lies on the line $j$, only values $\chi_{pq'}$ on the line $j-k$
contribute to the terms of order $k$ in \eqref{chistar}. In particular
the linear term in the expansion of ${\chi_{pq'}}^\star$ must vanish
for all $p$ and $q$, if $\chi$ is IGNR, i.e.\ for $q$ and $p$ at least 1,
\begin{equation}\label{linearterm}
    p B \chi_{(p-1)q'} +q \bar{B} \chi_{p(q-1)'}=0.
\end{equation}
This is therefore a necessary condition for IGNR. It will be used below
to obtain sufficient conditions.

The following lemma excludes the case $m=1=n$. In this case
we will have (by the argument in the proof below) $\chi_{00'}=0$ and
then  \eqref{linearterm} for $p+q=1$ would ensure IGNR.

\begin{lemma} If $\chi_{ab'}$ is IGNR and $m \geq 2$ then all values
  on the lines $k$, $0\leq k \leq(m-1)$ are zero.
\end{lemma}

\noindent {\bf Proof:}\\
We have already dealt with $n=0$ for all $m$.
For $a=m$, $b=0$, \eqref{chistar} gives the modified form of
\eqref{mrow} in which an index $0'$ is added to each occurrence of
$\chi$, i.e.\ 
$$ {\chi_{m0'}}^\star= \sum_{r=0}^m
\left( \begin{array}{c}m\\r\end{array}\right) B^r \chi_{(m-r)0'}.$$ By
  the same argument as for \eqref{mrow} $\chi_{a0'}=0$ for $0 \leq a <
  m$. (Similarly if $n\geq 1$ $\chi_{0b'}=0$ for $0 \leq b <n$.) Thus
  $\chi_{ab'}=0$ for $a+b=0$.
  If $m\geq 2$ and $n\geq 1$, $\chi_{10'}=0$ and
  \eqref{linearterm} implies $\chi_{01'}=0$ thus proving the result
  for $a+b=1$.

For $2\leq k\leq m-1$ we can prove $\chi_{ab'}=0$ on the line $k$ by
induction.  Suppose $\chi_{ab'}=0$ on all lines $i$, $i\leq j$ and
consider \eqref{chistar} for $k=j+2\leq m+1$. Terms of order 2 or more
there vanish by the induction hypothesis and we have only
\eqref{linearterm} with $a+b=j+1<m$ as a condition for IGNR.  For
$a=j$, $b=1$ this implies $\chi_{j1'}=0$ (since $\chi_{(j+1)0'}=0$)
and incrementing $b$ (and decrementing $a$) along the line $j+1$ we
similarly obtain $\chi_{ab'}=0$ for all values on the line $j+1$.

The induction step fails at $j+1=m$ because we have no equation giving
$\chi_{m0'}$. We now have to consider IGNR for values on the lines $k \in
(m,\,m+n)$.

\begin{lemma} If \eqref{linearterm} is satisfied for all values on the
  lines  $a+b-1=k \in (m,\,m+n-1)$, then $\chi_{ab'}$ is IGNR.
\end{lemma}

\noindent{\bf Proof:}\\ For an entry $\chi_{ab}$ on a line $\ell>m$
the term homogeneous in $B$ and $\bar{B}$ of degree $k > m-\ell$ is
zero as a consequence of Lemma 1.  For the terms homogeneous of degree
$k \leq m-\ell$ in $\chi_{ab'}$, \eqref{chistar} gives the value
\begin{equation}\label{chistark}
  {\chi_{ab'}^{(k)}}^\star =
    \sum_{s=0}^k   \left( \begin{array}{c}a\\k-s\end{array}\right)
    \left( \begin{array}{c}b\\s\end{array}\right)
      B^{(k-s)}\bar{B}^{s} \chi_{(a-k+s)(b-s)'}.
\end{equation}
      A direct check shows that this is equal to
      \begin{eqnarray}\label{kexpanded}
        &&\frac{1}{k}\{\sum_{s=0}^{(k-1)}
        \left( \begin{array}{c}a\\k-s-1\end{array}\right)
        \left( \begin{array}{c}b\\s\end{array}\right)
      B^{(k-s-1)}\bar{B}^{s}\nonumber \\
      \label{expr1}
      &&    \phantom{kB}  [(a-k+1+s)B\chi_{(a-k+s)(b-s)'}
        +(b-s)\bar{B}\chi_{(a-k+s+1)(b-s-1)'}]
      \}.
      \end{eqnarray}
Thus  ${\chi_{ab'}^{(k)}}^\star$ is zero if \eqref{linearterm} holds
for all pairs $(p,\,q)$ on the line $p+q=a+b-k$. One may note that
apart from the factor $1/k$ the
coefficients of the brackets containing the $\chi$ terms in \eqref{expr1} are
those which appear in the expansion of $\chi_{ab'}^{(k-1)}$, suggesting
a recursive or iterative proof could also be given.

The equality between \eqref{chistark} and \eqref{kexpanded} provides
in a general form the relationships that \citeasnoun{PolSkedIn00} overlooked. 

Combining the two preceding lemmas and the remark concerning $m=n=1$
we see that the following holds.
\begin{theorem}
  A totally symmetrized spinor $\chi_{ab'}$ with valence $(m,\,n)$
  such that $m\geq 1$ and $m \geq n$ is invariant under a
  one-dimensional group of null rotations which preserves the $o$
  basis spinor and has parameter $B$ in
  \eqref{nulrot}  if and only if
  \begin{eqnarray}
    \label{zerocond}
    \chi_{ab'}=& 0 & \forall~ {\rm pairs}~a+b\leq m-1,\\
    \label{linearcond}
    0=& (a+1) B \chi_{ab'} +b \bar{B} \chi_{(a+1)(b-1)'} &\forall~
    {\rm pairs}~a+b\in (m,\, m+n-1).
  \end{eqnarray}
\end{theorem}

One may note that, in accordance with the result stated by
\citeasnoun{PolSkedIn00}, the conditions
\eqref{zerocond}--\eqref{linearcond} imply that a two-parameter group
of null rotations is only possible if all $\chi_{ab'}=0$ except
$\chi_{mn'}$.

The conditions \eqref{zerocond}--\eqref{linearcond} are relatively
easy to check, provided one has aligned the frame so that $\npk$ is
the vector preserved by the null rotation. Nonzero Weyl tensors of Petrov
type N can be IGNR, and then this means taking a frame in which only
$\Psi_4\neq 0$ (or taking the alternative choice with only $\Psi_0\neq
0$, with consequent swapped values as above). This is generally easy
to achieve.

In conformally flat spacetimes, for the Segre types with non-zero
Ricci tensor which admit a null rotation isotropy the canonical forms
in Table 3 of \citeasnoun{PolSkedIn00} have $\Phi_{22'}\neq 0$, and for
types [(1(1,2)], {}[(1,3)], and [(11,2)] these are also the forms
satisfying the tests above. However, for Segre type [11(1,1)] the
canonical form in which the above conditions for null rotation
isotropy (about both $\npk$ and $\npl$) are manifest is not the one in
that Table. Instead one has to take a form in which only $\Phi_{11'}$
and $\Phi_{02'}$ are non-zero and $2|\Phi_{11'}| = |\Phi_{02'}|$: in
that form the conditions are satisfied for both null rotations with
the same parameter sets for $B$ and $E$. For a complete procedure for
checking for null rotation invariance in conformally flat spacetime
one needs a process for identifying Segre types that may have IGNR,
and then aligning the frame to the appropriate one of these canonical forms.

Assuming one has the frame well aligned, checking \eqref{zerocond} is
straightforward. Checking \eqref{linearcond} for a one-parameter group
with $B=|B|e^{i\theta}$ could be done in at least two ways. One could
first apply a rotation to obtain a frame in which $B$ became pure
imaginary (or, if preferred, real), obtaining the required rotation
angle from one of the relations \eqref{linearcond}. Since the
requirements imply that values along a line are simple multiples of
one another, this should not entail nontrivial division of polynomials
or other functions.  After applying the rotation found, one could then
check that the values on each line $a+b=k$, for $m+n-1 \geq k\geq m$,
satisfy \eqref{linearcond}.

An alternative is to take any nonzero pair of values of $\chi$ satisfying
\eqref{linearterm} and then for any other pair $(p,\,q)$ construct a
quadratic in components which will vanish if the $(p,\,q)$ entries
also satisfy \eqref{linearcond}. For example if the line $k=m+n-1$
contains nonzero terms one could check that
$$mq\chi_{p(q-1)'}\chi_{(m-1)n'} = np\chi_{(p-1)q'}\chi_{m(n-1)'}$$
for all other $p+q=k$,  $m+n-1 > k\geq m$.

In both ways of proceeding, one has to allow for the possibility that
for some metrics not all lines with $k>m$ are populated with nonzero
expressions for the components\footnote{One then has to bear in mind,
  if using computer algebra software, that the software may be unable
  to determine whether or not a particular expression is equivalent to
  zero.}. This was pointed out to me by Jan {\AA}man. An example is
provided by $\nabla^4\Psi_{ab'}$ for the metric (12.36) in
\citeasnoun{SteKraMac03} where the lines $k=9$ and 11 contain only zeroes,
although there are nonzero expressions in lines $k=8$, 10 and 12 (here
$m=8$ and $n=4$).

To carry out those checks one first has to locate the nonzero entries
in $\chi_{ab'}$. Jan {\AA}man has proposed doing so by finding the
nonzero entry in $\chi_{ab'}$ which is first in numerical order. He is
preparing a CLASSI module for making the required checks beginning with
that strategy.

In applying the above theorem when checking for null rotation isotropy
while classifying a spacetime metric, it may be useful to note that if
there is such isotropy then the same isotropy group must apply to all
the symmetrized spinors studied. So one could use the ratios found in
one such spinor to construct the quadratic test for another such
spinor.

\section*{Acknowledgement} I am grateful to Jan {\AA}man for comments
on previous drafts of this paper, some of which are incorporated
above, and for calculations of examples, as well as for the software
CLASSI used in investigating this issue, and to an anonymous referee
for prompting a clarification of the exposition.

\section*{References}

\end{document}